\documentclass[conference]{IEEEtran}

\usepackage{cite}
\usepackage{graphicx, epstopdf}
\usepackage{pgfplots}
\usepackage[cmex10]{amsmath}
\usepackage[caption=false,font=footnotesize]{subfig}
\usepackage{listings}
\usepackage{pgfplotstable}
\usepackage{url}
\begin{document}
\title{The Role of Big Data on Smart Grid Transition}

\author{\IEEEauthorblockN{The-Hien Dang-Ha\IEEEauthorrefmark{1},
		Roland Olsson\IEEEauthorrefmark{2} and
		Hao Wang\IEEEauthorrefmark{3} 
		}
		
		\IEEEauthorblockA{\IEEEauthorrefmark{1}\small Department of Informatics,
     	University of Oslo, Norway, Email: hthdang@student.matnat.uio.no}
		\IEEEauthorblockA{\IEEEauthorrefmark{2}\small Faculty of Computer Sciences
		Ostfold University College, Ostfold, Norway, Email: roland.olsson@hiof.no}
		\IEEEauthorblockA{\IEEEauthorrefmark{3}\small Faculty of Engineering and Natural Sciences, Aalesund University College, Email: hawa@hials.no}
}

\maketitle
\begin{abstract}
Despite being popularly referred to as the ultimate solution for all problems of our current electric power system, smart grid is still a growing and unstable concept. It is usually considered as a set of advanced features powered by promising technological solutions. In this paper, we describe smart grid as a socio-technical transition and illustrate the evolutionary path on which a smart grid can be realized. Through this conceptual lens, we revealed the role of big data, and how it can fuel the organic growth of smart grid. We also provided a rough estimate of how much data will be potentially generated from different data sources, which helps clarify the big data challenge during the evolutionary process.
\end{abstract}
\section{Introduction}
Although the term ``smart grid'' has become a popular concept, which is usually used to refer to the ultimate solution for all problems of our electric power system, it does not yet have a unique definition. The two most widely accepted and early definitions of the smart grid are the following:

\begin{itemize}
	\item \emph{US Department of Energy\cite{USEnergyDefinition}: } ``A smart grid uses digital technology to improve reliability, security, and efficiency (both economic and	energy) of the electrical system from large generation, through the delivery systems to electricity consumers and a growing number of distributed-generation and storage resources.''
	\item \emph{European Technology Platform on Smart Grid\cite{EuropeDefinition}: } ``A smart grid is an electricity network that can intelligently integrate the actions of all users connected to it--generators, consumers, and those that do both--in order to efficiently deliver sustainable, economic and secure electricity supplies.''
\end{itemize}

Obviously, these definitions try to deliver the meaning of smart grid in term of its expected features, properties, and abilities. This descriptive approach is not neutral nor self-evident, since it heavily depends on which actors or parts of the system are being focused. While the US definition pays more attention on the overall system and promising technological solutions, the European definition tends to target the end users. In fact, the smart grid actually means many different things, as generally listed in \cite{Morgan2009}. 

There was a debate at the ``Groningen Energy Summer School 2015'' organized by the Energy Academy Europe\cite{SummerSchool}, when the participants started discussing the definition of smart grid. We were wondering why it was named as it is instead of ``the future grid'', ``the next generation grid'', or ``the grid 2.0''. Since it was a multidisciplinary seminars and people were looking at the smart grid under many different perspectives, we could not come up with the final satisfying answer. However, people agreed that although the boundary of smart grid keeps expanding, incorporating more and more social and technological solutions, the core concept of smart grid relies on advanced digital information and communication technologies.

It is safe to say that the majority of smart grid features are powered by the massive amount of data generated by uncountable number of Internet-connected automated/controllable/programmable equipments. In fact, the smart grid can be seen as a huge sensor network, constantly collecting data from many different sources such as various sensors, smart meters, smart appliances, electrical vehicles, or even weather stations. Then, this immense amount of data can be transformed into actionable insights by applying high-volume data management and advanced analytics (i.e. Big Data) \cite{is2012managing}. These insights may help improve the efficiency, reliability and sustainability of the power grid through different use cases such as outage detection and restoration, preventive maintenance, shaping customer usage patterns, demand response, distribution system automation, or emission control. In other words, we believe that the smartness of the smart grid concept will be brought by the power of ``big data'' analytics.

Although there are a lot of publications and researching efforts on how we can overcome the big data challenges in some specific smart grid use cases \cite{bera2015cloud, diamantoulakis2015big, fan2012power, rusitschka2010smart, simmhan2013cloud}, there are very limited works on analyzing the smart grid transition as a whole to identify the role of big data during the process. In this paper, we bridge the gap by considering smart grid evolution as a socio-technical transition. Through a multi-level perspective, we analyze the path on which the smart grid can be realized, and identify the role and challenges of big data along the way. The paper is structured as following: Section \ref{sec:currentGrid} describes the current European power grid and illustrates the organic growth of smart grid; Section \ref{sec:socialtechnical} describes the smart grid evolution as a socio-technical transition in order to reveal a better overview of what is happening in the electricity infrastructure; Section \ref{sec:bigdata} clarifies the role of big data and provides a rough estimate of how big the data we must handle in order to build the according smart grid features; Section \ref{sec:conclusion} closes the paper by summarizing its main points.
\section{The Smart Grid Evolution}
\label{sec:currentGrid}
The electricity networks are arguably the biggest machines that humans have ever built. It comprises both the transmission infrastructure and the distribution infrastructure. The former operates at high voltage lines, moving large blocks of power between regions or countries, while the later operates at medium and low voltage lines, distributing and delivering the power to final customers. The boundary between transmission and distribution (i.e. the operated voltage levels) varies very much from country to country. In Europe, some distribution system operators (DSOs) also partly operates high voltage networks (where high voltage is more than 50 kV).

\subsection{How Smart Is the Current Transmission System?}
\begin{figure}
	\centering
	\includegraphics[trim=2.1cm 7.5cm 0.5cm 4cm, width=1\linewidth]{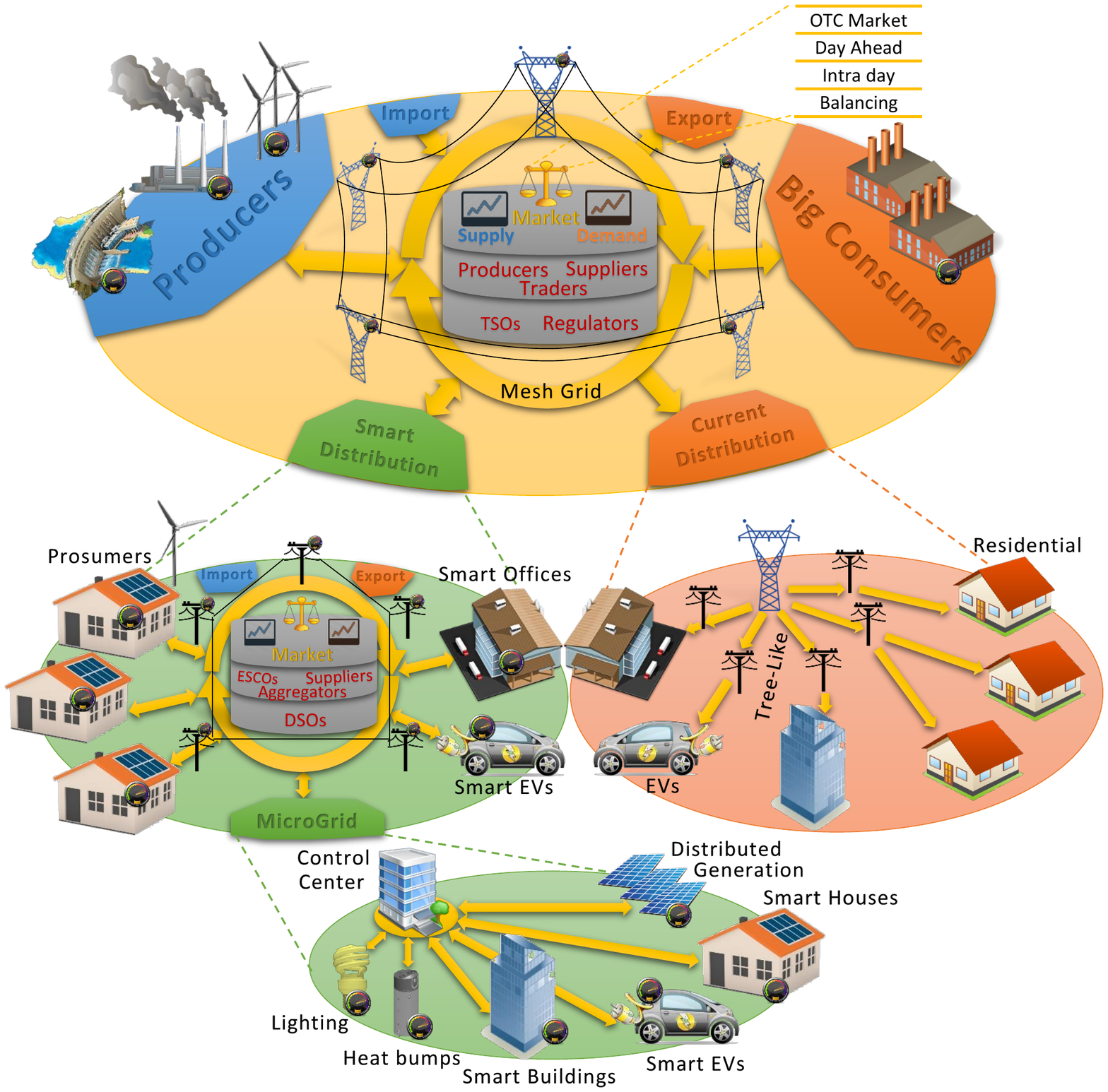}
	\caption{The electricity grid on the evolutionary path. \emph{Top:} The current transmission infrastructure; \emph{Middle:} the current and smart distribution grid operating side by side; \emph{Bottom: } the plug-and-play microgrid. The yellow arrows show the possible power flows, where two-way arrows indicate that the components can be controlled/scheduled to help balance the system. The meter icons represent all kinds of smart equipments belong to the advanced metering infrastructure}
	\label{fig:Grid}
\end{figure}

The power grid is sometimes described by the media as a hundreds-year old system, employing outdated technologies and highly vulnerable. The fact that we are trying to build the smart grid does not mean that the current grid is being passively or manually operated. The transmission infrastructure is already very smart with a large number of sensors and automated equipments. It is being monitored and regulated every second (or even more) by exceptionally complicated algorithms, keeping it stable all day every day. The high-voltage transmission lines are laid out as a ``mesh grid'', where the lost of one or even some units will not be able to kill the system. The transmission system operators (TSOs) can monitor, analyze, and adjust the power flows remotely from their control centers. The whole super complex infrastructure is designed and operated to serve a single purpose: \emph{balancing between electricity demand and supply}, as illustrated in the Figure \ref{fig:Grid}

This is a challenging task, since so far there is no technology that allows us to store huge amounts of energy effectively at a feasible cost. Besides, the consumer demand is an uncertain measure and can only be forecasted to a certain level of accuracy at specific horizons. Moreover, there are always unexpected events occurring on the grid. Therefore, matching between demand and supply requires a whole lot of planning at various horizons from years to seconds.

In the current grid, the balancing area is usually very large, which can even stretch over multiple countries. As depicted in Figure \ref{fig:Grid}, this balancing task is done by the wholesale electricity market, which usually involves
the following participants with different roles and responsibilities:
\begin{itemize}
	\item \emph{Producers: } they are responsible for power production and their primary objective is to maximize the profits and efficiency of their assets. Their generating capacity can also be traded to the TSOs, so it can be used as reserve capacity.
	\item \emph{Big consumers: } they are large-scale end-users, which can directly participate in the OTC or exchange markets to limit their exposure to the fluctuations in electricity prices. Sometimes, they also trade the right to reduce or shift their consumption to the TSOs, so that it can be used as a reserve capacity for balancing purpose. Other small-scale end-users are served by suppliers (i.e. retailers) through different types of contract such as fixed price or market price contract.
	\item \emph{Suppliers: } They either buy power directly from producers through OTC market, or indirectly through exchange market. Then, they basically resell the power to small and medium-sized consumers and households.
	\item \emph{Traders: } They can buy power from producers, retailers, or other traders, and sell it to another retailer. They help to create different routes on which the power can be exchanged from the producers to the end-users, and earn profit on price differences and volatility.
	\item \emph{Transmission System Operators (TSOs): } They are non-commercial neutral organizations, responsible for handling non-predictable imbalances and unexpected events on their area during real-time operations. They manage real-time balancing market and maintain the reserve and incidental emergency capacity: all the tools that can be used to balance any unexpected mismatch between supply and demand. They also determine the constraints in the transmission systems, ensure the system's long-term ability to meet the future transmission demands, and set the transmission tariffs. 
	\item \emph{Regulators: } They determine guidelines and bylaws for cost recovery through network tariffs, and resolve disputes.
\end{itemize}

 Because the demand and supply must be balanced at all times, and the price does not work fast enough to constantly clear the market, we need various markets for different time horizons so that participants can plan their demand or supply in advance. In this paper, we briefly explain the Nordic power market called Nord Pool\cite{spot2009nordic, Matias2012}, which is the largest European electrical energy market. These four following markets are currently available in Nord Pool:
\begin{itemize}
	\item \emph{Bilateral (OTC) Market: } Participants can trade electricity or rights to generating capacity for a specified period of time in the future such as next month, quarter, or year. The Nord Pool platform provides clearing service and plans to offer a reporting solution for OTC trades by the April 2016.
	
	\item \emph{Day-ahead Exchange Market (Elspot): } This is the main arena for trading power in the Nordic and Baltic region, accounted for majority of the volume handled by Nord Pool. Participants, according to their estimated demand or supply, place bids for a given quantity of power, at a given price and for a given hour on the following day. The auction closes at 12:00 the day prior to delivery. Based on the bids and transmission capacity of each bidding area, the price for each region is calculated using a common European advanced algorithm, which basically set where the demand and supply curves meet.
	
	\item \emph{Intra-day Exchange Market (Elbas): } Since the bids in day-ahead market have to be placed before 12:00 the day prior to delivery, incidents may take place between the gap. Intra-day market helps secure the balance between supply and demand by allowing seller and buyer continuously place bids until one hour before delivery. Prices are set based on first-come, first-serve principle.
	
	\item \emph{Balancing (Real-Time) Market:} This is operated by the TSOs to collect reserve capacity for any unexpected imbalance situation. Both the supply and demand side can offer capacity to the TSOs. It is further divided into three separate sub-markets based on the speed at which the reserves can be activated: 
		\begin{itemize}
			\item The primary market: reserves that can be activated immediately.
			\item The secondary market: reserves that have slightly longer response time.
			\item The tertiary market: reserves that have a manual response time of 15 minutes
		\end{itemize} 
	The first two markets are formed by agreements between TSOs and producers or big consumers through OTC marketplace, while the last market is operated at Nord Pool exchange marketplace.
\end{itemize}

As you can see, balancing between demand and supply is a very challenging task. So the transmission infrastructure has evolved into a proactive automated system, employing complex market-based control mechanism. Thanks to the high ramp rate and controllable behaviors of the traditional large-scale generators (e.g. hydro, thermal power, or gas turbine plants...), TSOs have the flexibility to adjust and keep the generation equal to the demand in real-time. In some cases, TSOs can also have the flexibility to shift or reduce the power consumption of some large industrial factories through contractual agreements with them. This demand side management solution was actually employed by the TSOs long before the smart grid idea was popularized.
 
 \subsection{How Smart Is the Current Distribution System?}
 As opposed to the transmission grid, most of the current distribution systems are made up of simple ``tree-like'' structures called ``distribution feeders'' \cite{Morgan2009}, as shown in Figure \ref{fig:Grid}. A distribution feeder typically has a single root point where the power flows in, and many branches where the power flows out to the customers. If something happens on or below a feeder (e.g. a tree falls on a power line, or a lightning  strikes a transformer) the entire feeder may be shut down automatically by circuit breakers. Surprisingly, typical distribution system operators (DSOs) are not able to detect these events until some customer has to call them. Indeed, the traditional role of DSOs is to ``building and connecting'', with minimal effort paid on the network management. They collect very sparse information and operate the low-voltage distribution network with very limited automation \cite{rusitschka2010smart}. This explains why nearly 90\% of all power outages and disturbances occur at the distribution network \cite{Farhangi2010}.

\subsection{The Path of the Smart Grid}
With the increasing demands for electricity, the need for integrating distributed renewable generation and carbon footprint reduction, the popularity of electric vehicles (EVs), and the active participation of end users; the power grid is facing big challenges as well as great opportunities to evolve. 

Besides the traditional centralized flexibility sources such as large-scale generators (and sometimes big consumers), the flexibility now can be collected from end-users through smart appliances, micro combined heat and power ($\mu$CHP), heat bumps, or electric vehicles (EVs). These distributed flexibility sources can be used to increase the grid capacity, efficiency, reliability, and security. The whole process is usually referred to as ``Demand Response'' (DR) or ``Demand Side Management'', which is one of the core concepts of smart grid.

Moreover, under the pressure of climate change and the depletion of fossil fuels, more renewables need to be included in the generation mix. Within this renewables mix, distributed renewable generation is expected to take a major part, as the efficiency and cost-effectiveness of wind turbines and rooftop solar panels are increasing gradually. However, most of these renewable sources are intermittent and depend heavily on weather condition. Without the extra flexibility collected from DR program, the cost of accommodating these high variable generation sources will be immense.

As one can readily see, the point of departure for the smart grid is placed at the distribution systems. They are the roots of most of the traditional as well as the smart grid issues. To be able to incorporate all of these components, the first step is that the DSOs must install an advanced metering infrastructure (AMI), which provides two-way communication system to the meter. The DSOs, through AMI, can get instantaneous data about individual or aggregated demand and estimate the state of the network. They can impose consumption capacity, or deploy DR strategy through direct control or dynamic pricing. Although the hardware and communication costs of the AMI equipments have become much slower than before, it is still too high for DSOs to install AMI on the entire distribution systems. Instead, there will be an organic growth of new distribution systems, which includes forward-looking technologies but fully compatible with the existing legacy system.

H. Farhangi (2010) \cite{Farhangi2010}, in his famous paper called ``The path of the smart grid''\footnote{This paper has received more than $1000$ citations from 2010}, has described how the grid will evolve into a smart grid. He confirmed that the smart grid will not replace but coexist with the existing power grid. It will be a complement to the present grid with add-on features, capabilities, and capacities. This process will start at distribution level, as visualized in Figure \ref{fig:Grid}, where smart and old distribution grids operate side by side. A smart distribution system, depending on its geography, the diversity of load and energy sources, and the level of investment, could be built to support different functionalities. In Figure \ref{fig:Grid}, the smart distribution network is being operated based on a market-based control mechanism. Flexibilities are collected from smart EVs, smart offices, prosumers, attached microgrids, etc., and traded on the retail flexibility market. \emph{Suppliers} are responsible for contracting with the flexibility owners, while the \emph{aggregators} buy it from suppliers or end-users, accumulate and trade it on the market. This flexibility can be used by the DSOs to resolve congestion points or reduce load peaks. It can also be imported or exported from and to the wholesale market. Moreover, thanks to the smart meter and control units attached to substations, feeders, and transformers, DSOs are able to implement distribution automation. The electricity network has also evolved into a mesh grid with redundant connections. In addition, the whole or part of a smart distribution grid can evolve into a (smart) microgrid, which is an interconnected network that can function whether they are connected to or separate from the main grid (grid-tied and islanded modes)\cite{Farhangi2010}.

The smart grid evolutionary path will be further investigated in the next two following sections, where the concept of socio-technical transition is introduced, and the role of big data in nurturing the organic growth of smart grid is revealed.

\section{Smart Grid as a Socio-Technical Transition}
\label{sec:socialtechnical}
When we talk about smart grid, we often think about a set of new technologies, such as advanced meter infrastructure, distributed renewable energy, or EVs. However, a broader and more accurate view can be obtained by looking at the electricity infrastructure as a socio-technical system, which consists of many components that are interdependent and have mutual relationships with each other. Introducing a new technology into the system will depend on and affect the whole environment around it. For example, if we would like to change the current vehicle type into electric one, besides the challenges in the electric vehicle technology itself, we must take into account all its surrounding components. The current automotive transport system is designed for and support the traditional combustion vehicles with all the fuel infrastructure; road infrastructure; production system; maintenance and distribution networks; regulations and policies; as well as many other components \cite{geels2002technological}. Even the driver practice that expect a vehicle with a long range and fast refueling is also a big barrier. The whole system need to be adjusted to make it compatible for the introduction of the new technology. That is one of the reasons why these changes take very long time, even though the required technology might be already available for years.

\begin{figure*}
	\centering
	\includegraphics[scale = 0.74, trim=0.5cm 17cm 1.5cm 2cm]{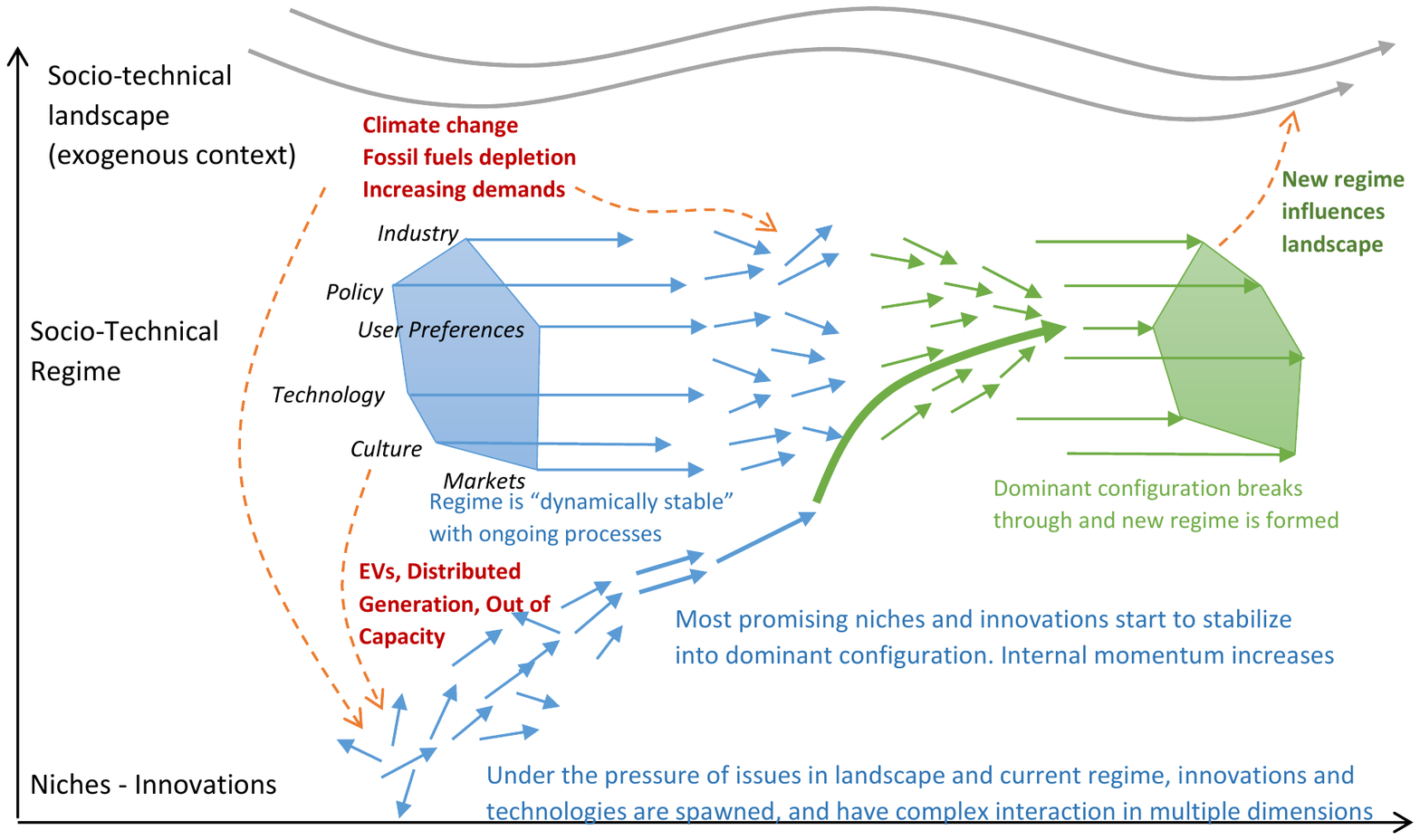}
	\caption{The multi-level perspective on smart grid transition (adapted from \cite{geels2007typology})}
	\label{fig:SocialTechnical}
\end{figure*}

The fundamental idea of the socio-technical transition concept is that the transition cannot be achieved by planning and exerting control. Instead, the system transforms itself into another ``dynamic stable'' configuration through self-organization and co-evolution processes, under the pressure of innovations, technologies, and other factors. A multi-level perspective on transitions has been introduced in \cite{geels2007typology}, which considers transitions as outcome of alignments between developments at multiple levels. The concept is adapted for smart grid and illustrated in Figure \ref{fig:SocialTechnical}.

The multi-level perspective divides a socio-technical system into three levels:
\begin{itemize}
	\item \emph{Socio-Technical Regime: } this is a dynamically stable configuration of the system. There are ongoing processes on different dimensions, which make it relatively stable within certain borders. It consists all the regulations, laws, infrastructures, practices, and organizations, etc. As mentioned above, it tends to resist to change, and normally takes many years until another configuration of the regime becomes dominant and a new regime is formed.
	\item \emph{Niches - Innovations: } this is the low-level place where niches and innovations are spawned. Niches are alternative solutions or alternative subsystems that could potentially replace certain components of the regime. They are small, incorporate different solutions and can change the way they function rapidly. At the beginning, there are many different niches and innovations, competing or supporting each other. After some time, the most promising innovations and niches become dominant, converge, and start to challenge the current regime. If the regime is not stable anymore due to forces from the landscape, the niches, and the regime itself, it will be replaced by the new dominant configuration.
	\item \emph{Landscape: } this is the exogenous environment in which the regime and niches exist. It includes external factors, such as public pressure on climate change or fossil fuel depletion. The niches and regime components will be affected and respond accordingly to the changes in the landscape.
\end{itemize}

This multi-level perspective supports the idea of the path of smart grid presented in the previous section. In this case, niches are under-construction smart distribution systems. They are basically pilots that utility companies use to test alternative technologies and solutions. If they generate good return on investment in a short enough period of time, the quality and quantity of these pilots will grow organically. In the beginning, these pilots are very different (e.g. use different solutions or innovations) and have complex interaction (e.g. competing, supporting...) in multiple dimensions. But later, the most promising pilots and innovations start to stabilize into dominant configuration. With the help from policy makers, marketers, media, etc., the current electricity infrastructure regime can be unstabilized and open for transition. The dominant configuration then can challenge and potentially replace the current regime.
\section{Big Data Role and Challenges}
\label{sec:bigdata}
As you can see, the smart grid transition requires many forces from different actors before the current regime can be transformed. However, one of the biggest challenges is how to quickly make enough profits from niche investments, so that the quality and quantity of them can grow organically. We believe that Big Data is the solution, and the major driving force behind the growth.

\begin{figure}
	\centering
	\includegraphics[trim=2cm 10cm 3cm 3.5cm, width=1\linewidth]{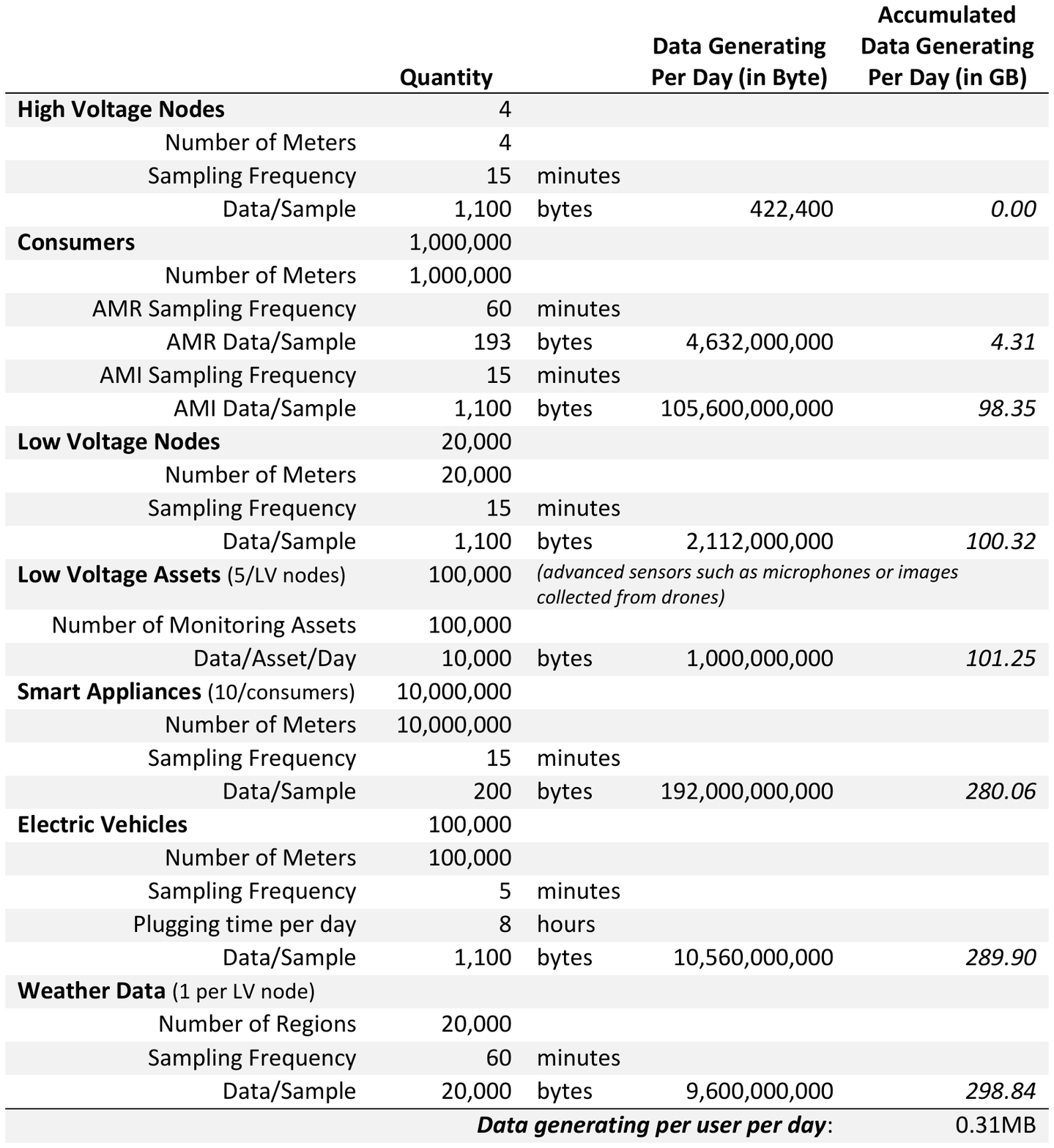}
	\caption{Data generating potential estimated for 1M end-users. The ratio between the high voltage nodes, low voltage nodes, and the end-users is equal to the current ratio in the Netherlands (provided in \cite{aiello2014smart}). The data size per sample and the sampling frequency are adapted from the study conducted in \cite{aiello2014smart}. Interestingly, the data generated by an end-user each day is comparable to that of an average Facebook user.}
	\label{tab:DataPotential}
\end{figure}

\begin{figure*}
	\centering
	\includegraphics[trim=2.5cm 18.5cm 0.5cm 3cm, width=1\linewidth]{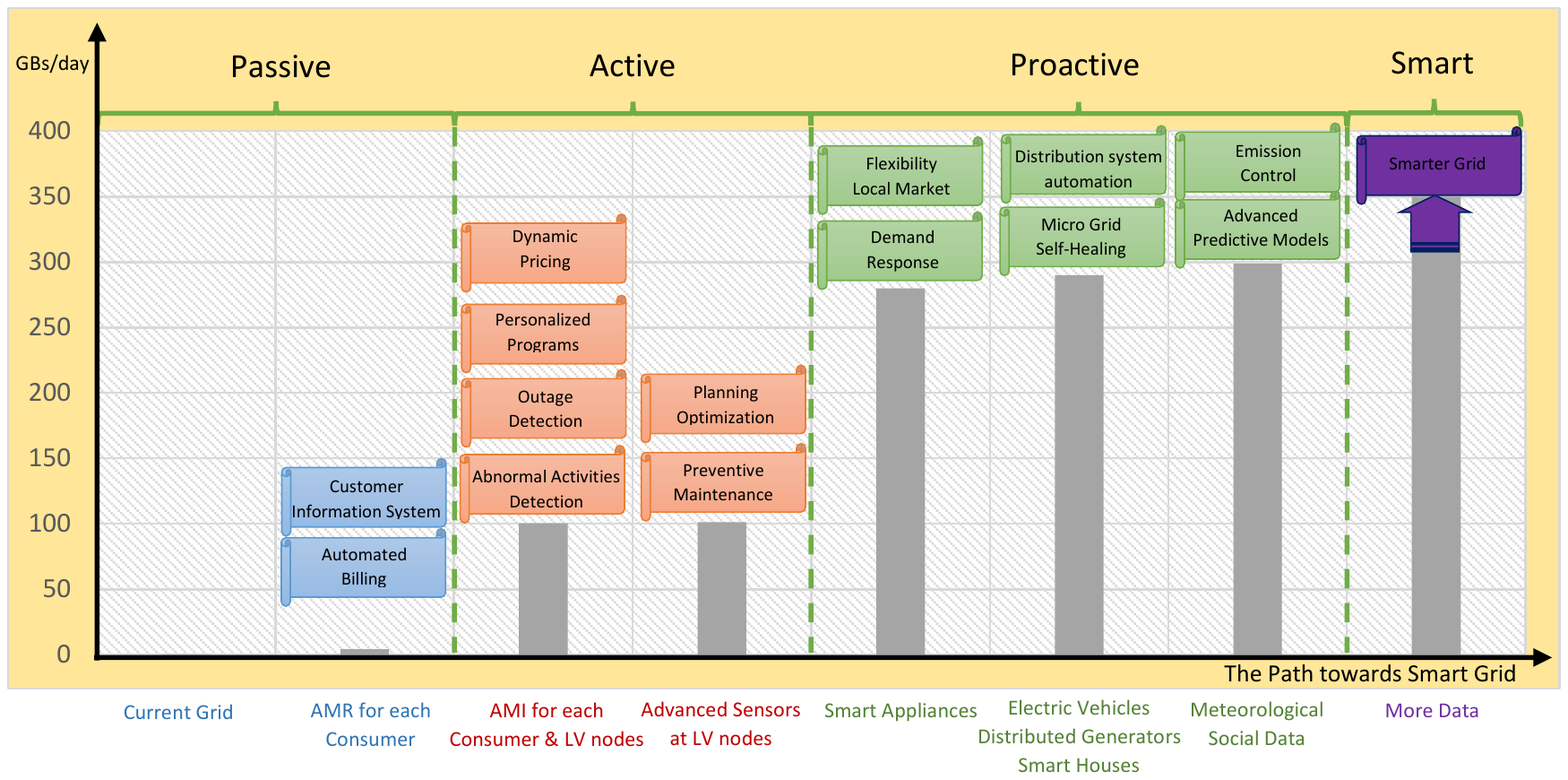}
	\caption{The evolutionary path towards smart grid, where each step involves connecting more data sources and creating more values by implementing corresponding big data applications. The data generating potential in each step is also illustrated to roughly show how big the data we must handle.}
	\label{fig:DataPotential}
\end{figure*}

As described above, the utility companies cannot simply deploy a massive advanced metering infrastructure, where every relevant component is connected by a corresponding smart equipment. This is just too costly to be feasible. Instead, this process should be done step by step, where the benefits generated from a certain step can partly cover the deployment cost of the next step. This is where the Big Data comes into the smart grid transition picture. The majority of smart grid use cases are characterized by the massive amount of data collected from all the connected smart equipments at near real-time. And big data is a set of tools that allow utility companies to handle and transform this high-volume high-velocity data into values.

Big Data Analytics tools such as data mining, pattern matching, and stream analytics, can reveal hidden models or correlations from the ocean of sensing data. These insights can be used to visualize, optimize, or automate the grid operations. Whenever a new data source is connected, certain big data applications can be implemented, which can add new features and create new values from the data. These values will fuel the organic growth of the smart grid, which leads to further investments and connected data sources. Figure \ref{fig:DataPotential} describes how a current passive distribution grid can become an active, then proactive, and finally a smart grid.

Based on the study conducted in \cite{aiello2014smart}, we adapted and normalized the figures to obtain a rough estimate of how much data will be potentially generated from each data sources. The overall calculation is given in Table \ref{tab:DataPotential}.

\section{Conclusion}
\label{sec:conclusion}
The paper summarizes the current European power grid situation, and illustrates how it can evolve into smart grid through an organically growing process. We further investigated the smart grid evolution as a socio-technical transition using the multi-level perspective. This conceptual lens allows us to reveal the big data role during the transition, and how it can fuel the organic growth of smart grid. We also did a quantitative study to estimate how much data will potentially be generated from different data sources, which help clarify the big data challenge on the path towards smart grid.
\bibliography{BigDataTowardsSmartGrid}
\bibliographystyle{IEEEtran}
\end{document}